# U-NET CNN BASED FOURIER PTYCHOGRAPHY

Yican Chen, Zhi Luo, Xia Wu, Huidong Yang, and Bo Huang*
College of Information Science and Technology, Jinan University, Guangzhou 510632, China

**Abstract**

Fourier ptychography is a recently explored imaging method for overcoming the diffraction limit of conventional cameras with applications in microscopy and yielding high-resolution images. In order to splice together low-resolution images taken under different illumination angles of coherent light source, an iterative phase retrieval algorithm is adopted. However, the reconstruction procedure is slow and needs a good many of overlap in the Fourier domain for the continuous recorded low-resolution images and is also worse under system aberrations such as noise or random update sequence. In this paper, we propose a new retrieval algorithm that is based on convolutional neural networks. Once well trained, our model can perform high-quality reconstruction rapidly by using the graphics processing unit. The experiments demonstrate that our model achieves better reconstruction results and is more robust under system aberrations.

## 1. Introduction

In imaging system, space-bandwidth product (SBP) is used to characterize the tradeoff between high resolution and large field of view and the total resolvable pixels [1]. Fourier ptychography (FP) [1] is an effective imaging technique that aims to tackle the physical limitation by capturing a sequence of SBP limited images and computationally combining them to recover a high resolution, large FOV image. It has applications in wide field, high resolution microscopy [1], microscopy biomedical imaging [2], long distance, sub-diffraction imaging [3] and other applications.

The FP method needs acquisition of many SBP limited images which are gained from varying illumination angles of coherent light source. Because conventional image sensors can measure only intensity information of light, there's a loss of phase information. As a result, iterative phase retrieval algorithm is applied to the recorded set of images to recover the phase information that is lost in the imaging process and thus reconstruct a high resolution, high field-of-view image. So far, most FP applications [3-6] utilize the Alternating Projection (AP) algorithm [7,8,10] to implement the reconstruction process.

However, the reconstruction quality of iterative phase retrieval algorithms degrades as the overlap between the successively captured images in the Fourier domain decreases [9]. Reconstruction under less overlap becomes even more challenging for incomplete amplitude and phase information. Hence, we need to capture more SBP limited images which not only spans the Fourier domain, but also fulfill the needed number of overlap radio. And also, the low reconstruction quality happens as the updating sequence for FP is random [10] and the robustness under system aberrations is poor [11]. In this paper, we focus on a deep learning-based algorithm in solving inverse retrieval problems for the task of phase retrieval in FP. Instead of a phase retrieval algorithm, we propose a Convolutional Neural Network (CNN) based method (FPNET) that directly restores the image in the spatial domain without explicitly recovering the phase information. CNNs have been demonstrated to provide superior performance to solve many challenging imaging problems, such as super-resolution [12,13], segmentation [14], deconvolution [15], holography [16], phase recovery [17], etc.

We show that FPNET obtains better reconstruction results in case of different overlapping frequency bands and case of noisy images and random sequence images. The remainder of the paper is organized as follows. In Section 2 we briefly introduce Fourier Ptychography and the AP algorithm, Section 3 contains our proposed method FPNET. In Section 4 we explain our results and experimental evaluation and Section 5 concludes the paper.

## 2. Principle of Fourier ptychographic microscopy

As a classic computational method, the FPM is mainly composed of the forward imaging model and the recover process. In the forward imaging model, the sample is illuminated by oblique plane waves from the LED matrix, and then the exiting waves are captured by the camera through the objective lens. By sequentially illuminating the different LEDs on the matrix, a series of low-resolution intensity images are obtained. In recover process, the high-resolution image is reconstructed with the low-resolution intensity images.

### 2.1. Forward imaging model

In the forward imaging procedure, we denote a thin specimen as its transmission function $o(r)$, where $\mathbf{r} = (x, y)$ represents the 2D spatial coordinates. Assuming that the LED is far enough away from the sample stage, the illumination wave is approximately oblique plane wave. For the $n_{th}$ LED, the wave vector can be expressed as

$$\mathbf{k}_n = \left(\frac{\sin \theta_{xn}}{\lambda}, \frac{\sin \theta_{yn}}{\lambda}\right) (n = 1,2,3, \cdots, N_{LED}), \tag{1}$$

where $(\theta_{xn}, \theta_{yn})$ represent the illumination angle for the $n_{th}$ LED and $\lambda$ is the wavelength [10]. Using an oblique plane waves with a wave vector $\mathbf{k}_n$ to illuminate the specimen is equivalent to shifting the specimen spectrum $O(\mathbf{k})$ to be centered around $\mathbf{k}_n$, expressed as $\mathcal{F}\{o(\mathbf{r})exp(i\mathbf{k}_n\mathbf{r})\} = O(\mathbf{k} - \mathbf{k}_n)$. The field is low-pass filtered by the objective lens with pupil function $C(\mathbf{k})$ when passing through the objective lens. The forward imaging model of FPM can be denoted as

$$I_{nc}(\mathbf{r}) = |g_{nc}(\mathbf{r})|^2 = |\mathcal{F}^{-1}\{C(\mathbf{k})O(\mathbf{k} - \mathbf{k}_n)\}|^2, \tag{2}$$

where $I_{nc}(\mathbf{r})$ expresses the intensity on sensor, $g_{nc}(\mathbf{r})$ expresses the complex field on sensor, $O(\mathbf{k} - \mathbf{k}_n)$ expresses the spectrum of the specimen illuminated by a plane wave with a wave vector $\mathbf{k}_n$, $\mathbf{k} = (k_x, k_y)$ represents the 2D frequency coordinates and $\mathcal{F}^{-1}$ denotes the inverse Fourier transform.

## 2.2. Recovering process

In recovering process, FP can synthesize those images with different spectral information and get an estimation of high-resolution complex field $o_e(\mathbf{r}) = \mathcal{F}^{-1}\{O_e(\mathbf{k})\}$. Generally, FP recovers the high-resolution image through the most classic reconstruction method, termed Alternate Projection (AP), which iteratively estimate the complex fields and update them with captured intensity images. One iteration can be denoted as

$$g_{ne}(\mathbf{r}) = \mathcal{F}^{-1}\{P(\mathbf{k})O_e(\mathbf{k} - \mathbf{k}_n)\} \tag{3}$$

and

$$P(\mathbf{k})O_e(\mathbf{k} - \mathbf{k}_n) = \mathcal{F}\left\{g_{ne}(\mathbf{r}) \frac{\sqrt{I_{nc}(\mathbf{r})}}{|g_{ne}(\mathbf{r})|}\right\}. \tag{4}$$

Equation (3) is to estimate the high-resolution images corresponding to each illumination and Eq. (4) is to update the high-resolution images using the captured low-resolution intensity images [1,10]. The premise of Eqs. (3) and (4) is that in practical applications, the sample should be strictly placed on the focal plane and the careful calibration is required [4]. Until the estimated spectrum converges, the manipulations in Eqs. (3) and (4) are repeated. The iteration starts from a random guess of $g_{ne}(\mathbf{r})$. At last, by inverse Fourier transforming the estimated spectrum $O_e(\mathbf{k})$ to $o_e(\mathbf{r})$, the high-resolution images are extracted from $o_e(\mathbf{r})$.

## 3. FPM reconstruction framework with FPNET

We propose a learning-based framework of reconstructing high-resolution image based on U-Net architecture [18], which is based on Convolutional Neural Networks. Our model records a non-linearly end-to-end mapping between the input of low intensity images and original high-resolution images. We also describe how to build datasets to train FPNET with FPM imaging model. The low-resolution images which are gained by using varying illumination angles of a coherent light, serve as input data of network.

### 3.1. Datasets building

We need a large training dataset to train the deep neural network and a small test dataset to test the network performance. Images in DIV_2K, Set14 and Set5 are used for generating datasets with simulations, as the strategies used in many deep learning image reconstruction projects. 800 images in DIV_2K are used as train sets and the latter two are used as test sets. The original images are converted to grayscale and cropped to w × h pixels, where w = h = 128 pixels. These images represent our ground truth data. With the FPM forward imaging model, we use these high-resolution complex fields to generate $N^2$(N=5) images with the 32 × 32 sized low-resolution intensity input of the model. The order of every degraded image sequence is randomly disrupted and then the images are concatenated into a 3D-cube of size 32×32×$5^2$. Using 800 images in

DIV_2K, approximately 30000 32 × 32 × 25 cubes were extracted to train the FPNET. Data preparation for training also includes resizing the cubes spatially to 128×128×25 with the bicubic interpolation algorithm, and performing channel wise rescaling to have values between 0 and 1. Similar manipulation is done for the ground truth high resolution images. We also create four separate training datasets with the radio (0%,18%,40%,65%) of overlapping frequency bands. Gaussian distributed noises with a mean of zero and a standard deviation of 0 to $3 \times 10^{-4}$ is added to the images in the dataset. An example of building the training dataset is shown in Fig. 1.

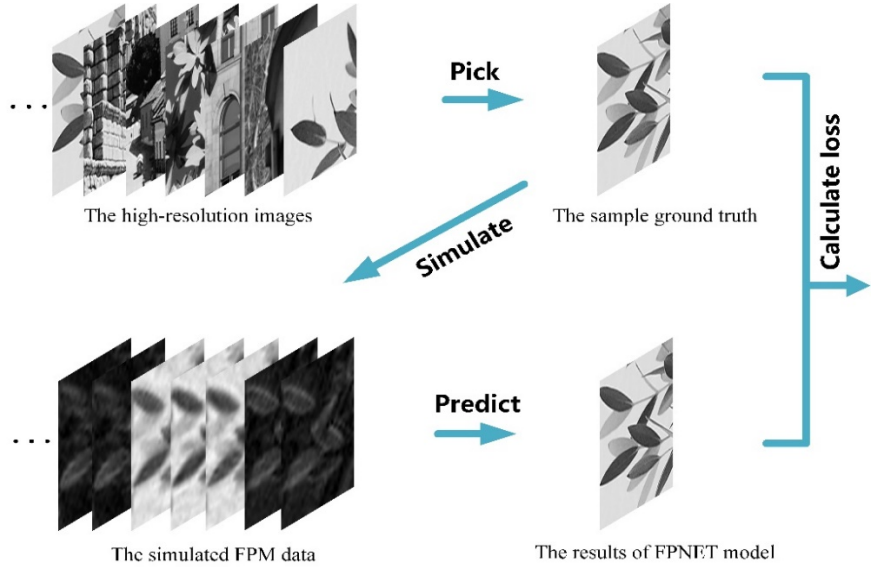

Fig.1. An example of how to build and use the training dataset

### 3.2. Reconstruction with FPNET

Unlike those iteration-based strategy [10,19], we use DNN to build model and train model to study the non-linear relations between input and output on a large-scale dataset. After training the model, the DNN can be used for prediction and quickly perform the reconstruction process.

We design the FPNET model based on the popular U-Net architecture. Fig. 2 shows a detailed diagram of the proposed DNN architecture. At the first layer a 3×3 convolution is used to receive the input low-resolution images. The middle layer stack consists of a contracting path (left side) and an expansive path (right side). The contracting path follows the typical architecture of a convolutional network. It consists of the repeated application of two 3×3 convolutions (padded convolutions), each followed by a rectified linear unit (ReLU) and a 3×3 down-convolution with stride 2 for down-sampling. Every step in the expansive path consists of an up-sampling of the feature map followed by a 2×2 up-convolution, a concatenation with the correspondingly feature maps from the contracting path, and two 3×3 convolutions, each followed by a ReLU. It is easy to see that a stack of two 3× 3 conv. layers (without spatial pooling in between) has an effective receptive field of 5×5. We decrease about 28% of the number of parameters: assuming that both the input and the output of a two-layer $3 \times 3$ convolution stack has C channels, the stack is parameterized by $2 \times 3^2 C^2 =$

$18C^2$ weights; at the same time, a single 5 × 5 conv layer would require $5^2C^2 = 25C^2$ parameters, i.e.. At the final layer a 1×1 convolution is used to map each 64-component feature vector to the desired number of classes. In total the network has 28 convolutional layers. There are two key advantages that recommend U-Net for our purpose.

1.Multi-resolution decomposition: The decoder uses a compression-expansion structure based on down-convolution and up-convolution. This means that given a fixed-size convolution kernel (3×3 in our example), the effective receptive field of the network increases as the input goes deeper into the network.

2.Local-global composition: In each resolution level, the outputs of the convolutional block in the contraction are directly connected and concatenated with the input of the convolutional block in the expansion. The skip connection, which combine deep, semantic, coarse-grained feature maps from the decoder sub-network with shallow, low-level, fine-grained feature maps from the encoder sub-network.

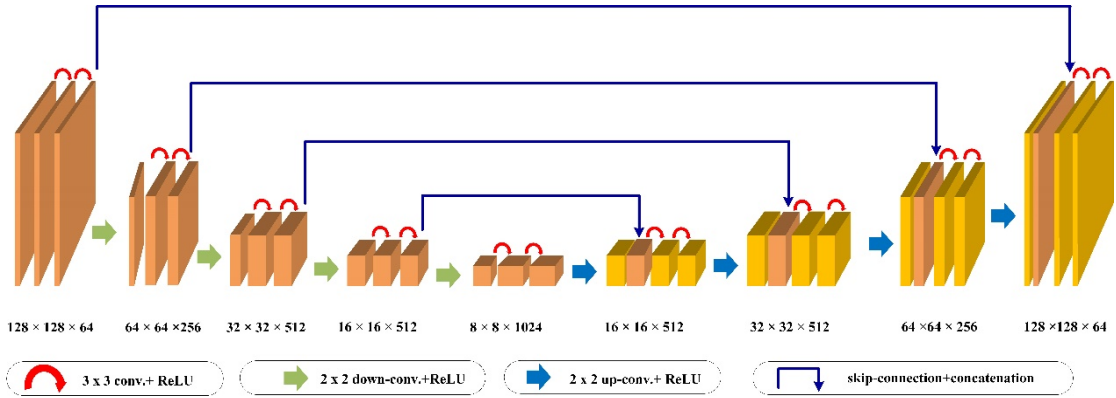

Fig. 2. Visual illustration of the proposed learning architecture (middle part) based on U-Net, indicating the type of layers, nodes in each layer, the numbers of feature map, etc.

## 4. Experiments

In this section, we show the results of our simulations and experiments. The FPNET is already pretrained with the simulation training dataset and tested with the test dataset. The hyperparameters of FPNET are carefully adjusted to reach the best performance in the training process. Specifically, we set the learning rate as $1 \times 10^{-4}$, the batch size as 32 and the patch size as 128. We train FPNET with 300,000 iterations. We use MAE (Mean Absolute Error) loss function to constrain the training. We perform FPNET and AP on the simulation test dataset under a series of overlapping radios. We also show the performance of AP and FPNET with a series of noise levels. To be clear, all results of FPNET in subsection 4.1 and 4.2 are trained respectively because their training datasets are completely different from each other. All methods are implemented with Python and run on a NVIDIA GTX 1080Ti graphics card.

## 4.1 Performance with random images sequence and different overlap

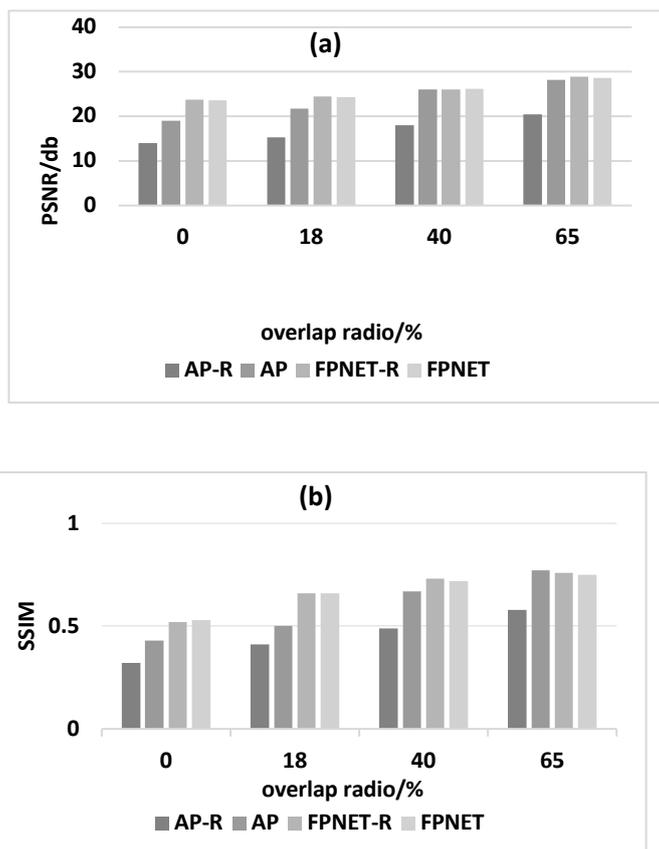

Fig.3. Evaluation of reconstruction results under a series of overlapping radios. AP-R and FPNET-R represents AP algorithm's and FPNET's test results of random image sequence respectively. (a) show the PSNR (Peak signal to noise ratio) of recovered images, (b) show the SSIM (structural similarity index) of recovered images.

| AP-R (0%) | AP (0%) | FPNET-R (0%) | AP-R (65%) | AP (65%) | FPNET-R (65%) |
|---|---|---|---|---|---|
| 14.62/0.366 | 15.78/0.387 | 23.53/0.584 | 21.14/0.645 | 25.65/0.789 | 28.65/0.716 |
| 15.05/0.412 | 17.07/0.412 | 24.07/0.568 | 23.22/0.578 | 25.27/0.758 | 26.38/0.703 |
| 16.58/0.313 | 18.19/0.431 | 24.83/0.538 | 22.85/0.602 | 27.99/0.729 | 31.28/0.741 |

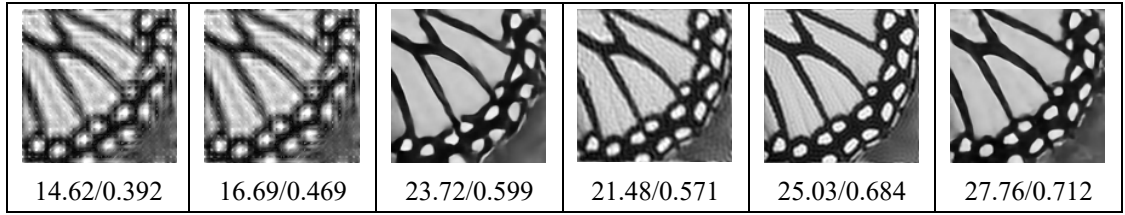

| 14.62/0.392 | 16.69/0.469 | 23.72/0.599 | 21.48/0.571 | 25.03/0.684 | 27.76/0.712 |

Fig.4.1. Example recovery results of AP and FPNET under 0% and 65% overlap radios on test dataset.

As discussed in Section 1, random updating sequence of intensity images and low overlap radio limit the reconstruction and temporal resolution of FPM. Reducing overlap radio of intensity images, the final resolution of reconstruction results will decrease, which is equivalent to add blurring to the final results. If high-resolution intensity can be reached with fewer overlap radio, the temporal resolution of FPM will greatly increase. If the updating sequence of intensity images are random, the recovery process is hard to converge to good results. To test the performance of FPNET with low overlap radio and random sort intensity images, we compare the reconstruction results of FPNET methods with intensity images with different overlap radio, random sort and the result of AP with the same images. The PSNR and SSIM of the reconstruction results are shown in Fig. 3 and Fig.4. The experiment demonstrates that FPNET can still reach a usable reconstruction result when greatly reducing the overlap radio of intensity images and disrupting intensity images' order. This characteristic of FPNET is very useful in conditions that requiring high temporal resolution. This experiment also proves that FPNET model is very helpful to overcome positional misalignment problem.

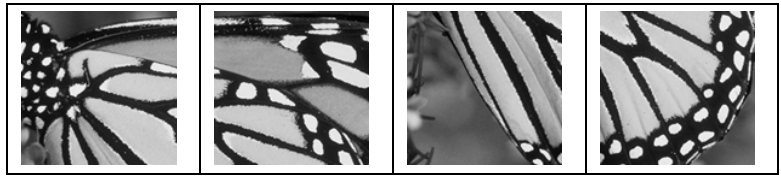

Fig.4.2. Ground truths' example of test dataset (correspond to Fig.5.1).

## 4.2．Performance under noise

| AP-1 | FPNET-1 | AP-2 | FPNET-2 | AP-3 | FPNET-3 |
|---|---|---|---|---|---|
| 23.86/0.621 | 27.93/0.739 | 21.67/0.557 | 26.84/0.712 | 20.11/0.516 | 26.08/0.709 |
| 23.71/0.636 | 27.51/0.721 | 21.48/0.569 | 27.05/0.695 | 20.63/0.521 | 26.44/0.689 |

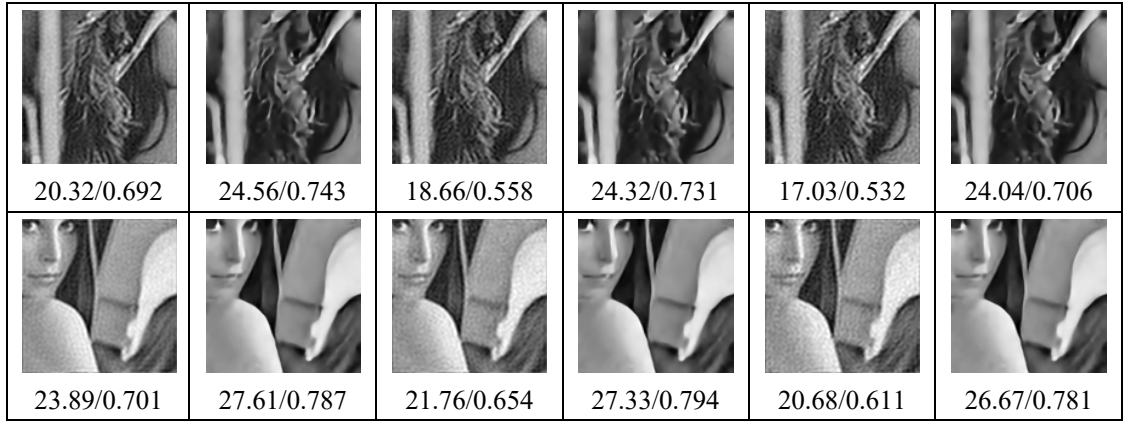

| | | | | | |
|---|---|---|---|---|---|
| 20.32/0.692 | 24.56/0.743 | 18.66/0.558 | 24.32/0.731 | 17.03/0.532 | 24.04/0.706 |
| 23.89/0.701 | 27.61/0.787 | 21.76/0.654 | 27.33/0.794 | 20.68/0.611 | 26.67/0.781 |

Fig.5.1. Example recovery results of AP and FPNET on the test dataset. Gaussian distribution noises with zero mean and standard deviation of $1\times10^{-4}$, $2\times10^{-4}$, $3\times10^{-4}$ are added on group 1,2 and 3.

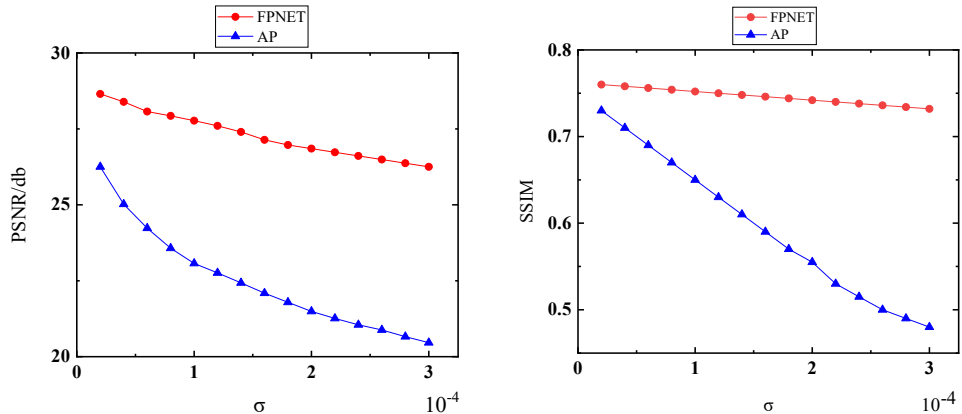

Fig.5.2. Evaluation of recovery results under a series of noise levels. The graph above show the PSNR and SSIM of recovery images under noise.

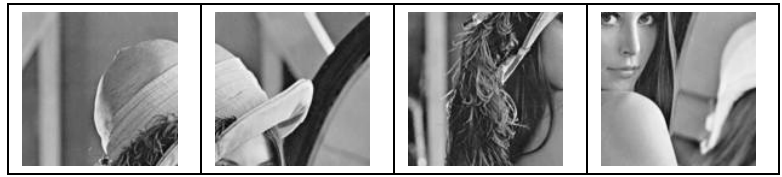

Fig.5.3. Ground truths' example of test dataset (correspond to Fig.6.1).

To evaluate the effectiveness of FPNET, we perform FPM reconstruction with FPNET and AP on the simulation test dataset and compare the results of these two methods. Under actual experimental conditions, the captured intensity images are contaminated by imaging noise which greatly affects the reconstruction results. To simulate the actual conditions, gaussian distribution noises with zero mean and standard deviation ranging from $2\times10^{-5}$ to $3\times10^{-4}$ are added on the test dataset respectively. The FPNET method is already trained on the training dataset before test. Fig.5.1 and Fig.5.3 show some example reconstruction results of these two methods and the ground truth. The PSNR and SSIM of these example images are also shown in the figure. Besides, Fig.5.2 show the PSNR and SSIM of AP and FPNET under different noise levels. As the plots and example results show, FPNET's results are more smoothly and have more details and fewer artifacts comparing with AP's. The increase in noise has little effect on the FPNET's result and the impact on the AP's results are more and more obvious. It can be concluded that FPNET performs better and more robust than AP under noise.

## 5. Conclusion

We introduced a reconstruction method for Fourier ptychography based on U-net structure. We demonstrated that it is superior to use the deep neural network to perform FPM recovery. Once our model is well trained, it can perform high-quality reconstruction rapidly by using the graphics processing unit. We also showed that for low-overlapped and non-overlapped Fourier sampling, FPNET performed significantly better than AP. Furthermore, FPNET is more robust than AP under system aberrations.

## 6. References.

[1]. Zheng, G., R. Horstmeyer and C. Yang, "Wide-field, high-resolution Fourier ptychographic microscopy," Nature Photonics, 2013. 7(9): p. 739-745.
[2]. Horstmeyer, R., et al., "Digital pathology with Fourier ptychography," Computerized Medical Imaging and Graphics, 2015. 42: p. 38-43.
[3]. Holloway, J., et al., "Toward Long-Distance Subdiffraction Imaging Using Coherent Camera Arrays," IEEE Transactions on Computational Imaging, 2016. 2(3): p. 251-265.
[4]. Sun, J., et al., "Efficient positional misalignment correction method for Fourier ptychographic microscopy," Biomedical Optics Express, 2016. 7(4): p. 1336.
[5]. Tian, L., et al., "Multiplexed coded illumination for Fourier Ptychography with an LED array microscope," Biomedical optics express, 2014. 5(7): p. 2376.
[6]. Zhang, J., et al., "Precise Brightfield Localization Alignment for Fourier Ptychographic Microscopy," IEEE Photonics Journal, 2018. 10(1): p. 1-13.
[7]. Fienup, J.R., "Reconstruction of an object from the modulus of its Fourier transform," Optics letters, 1978. 3(1): p. 27.
[8]. Gerchberg, R.W., "Super-resolution through Error Energy Reduction," Optica Acta: International Journal of Optics, 1974. 21(9): p. 709-720.
[9]. Jianliang Qian, Chao Yang, A Schirotzek, F Maia, and S Marchesini, "Efficient algorithms for ptychographic phase retrieval," Inverse Problems and Applications, Contemp. Math, vol. 615, pp. 261–280, 2014.
[10]. Guoan Zheng. "Fourier Ptychographic Imaging, 2053-2571. Morgan & Claypool Publishers," 2016. ISBN 978-1-6817-4273-1. doi: 10.1088/978-1-6817-4273-1. URL http://dx.doi.org/10.1088/978-1-6817-4273-1
[11]. Bian, Z., S. Dong and G. Zheng, "Adaptive system correction for robust Fourier ptychographic imaging," Optics Express, 2013. 21(26): p. 32400.
[12]. Dong, C., et al., "Image Super-Resolution Using Deep Convolutional Networks," IEEE Transactions on Pattern Analysis and Machine Intelligence, 2016. 38(2): p. 295-307.

[13]. Ledig, C., et al., "Photo-Realistic Single Image Super-Resolution Using a Generative Adversarial Network," in Proceedings of IEEE Conference on Computer Vision and Pattern Recognition (IEEE, 2017), pp. 105-144.

[14]. Jonathan Long, Evan Shelhamer, and Trevor Darrell, "Fully convolutional networks for semantic segmentation," in Proceedings of the IEEE Conference on Computer Vision and Pattern Recognition, 2015, pp. 3431–3440.

[15]. M. D. Zeiler, D. Krishnan, G. W. Taylor, and R. Fergus, "Deconvolutional networks," in Proceedings of IEEE Conference on Computer Vision and Pattern Recognition (IEEE, 2010), pp. 2528-2535.

[16]. Rivenson, Y., et al., "Phase recovery and holographic image reconstruction using deep learning in neural networks," Light: Science & Applications, 2018. 7(2): p. 17141-17141.

[17]. A. Sinha, J. Lee, S. Li, and G. Barbastathis, "Lensless computational imaging through deep learning," Optica 4(9), 1117-1125 (2017).

[18]. O. Ronneberger, P. Fischer, and T. Brox, "U-net: Convolutional networks for biomedical image segmentation," https://arxiv.org/abs/1505.04597.

[19]. Jiang, S., et al., "Solving Fourier ptychographic imaging problems via neural network modeling and TensorFlow," Biomed Opt Express, 2018. 9(7): p. 3306-3319.